# Highlights

**Unsteady Load Mitigation through Passive Pitch**

Yabin Liu,Riccardo Broglia,Anna M. Young,Edward D. McCarthy,Ignazio Maria Viola

- The dynamic process of a foil experiencing gusts is investigated by fluid-structure interaction simulations.
- The gusts considered are fast, high-amplitude variations of the free stream speed and direction.
- Passive pitching can mitigate gust-induced lift fluctuations by at least two-thirds.
- The optimal position of the passive pitching axis is in front and upstream of the foil.

# Unsteady Load Mitigation through Passive Pitch

Yabin Liu[a], Riccardo Broglia[b], Anna M. Young[c], Edward D. McCarthy[d] and Ignazio Maria Viola[a,*]

[a]*School of Engineering, Institute for Energy Systems, University of Edinburgh, Edinburgh EH9 3FB, United Kingdom*
[b]*CNR-INM, Institute of Marine Engineering, National Research Council, Rome 00128, Italy*
[c]*Department of Mechanical Engineering, University of Bath, Bath, BA2 7AY, United Kingdom*
[d]*School of Engineering, Institute for Materials and Processes, University of Edinburgh, Edinburgh, Robert Stevenson Road, EH9 3FB, Scotland, United Kingdom*

## ARTICLE INFO

*Keywords*:
Passive pitch
Unsteady loads
Gust mitigation
Fluid-structure interaction

## ABSTRACT

Mitigation of load fluctuations due to flow unsteadiness is critical in a broad range of applications, including wind/tidal turbines, and aerial/underwater vehicles. While the use of active control systems is an established practice in engineering, passive systems are not well understood, and the limits of their efficacy are yet to be ascertained. To this end, the present study aims to provide new insights into the effectiveness of passive pitching in the mitigation of lift fluctuations in the most demanding case of fast, high-amplitude variations of the free stream speed and direction. We perform fluid-structure interaction simulations of a two-dimensional free-to-pitch rigid foil. Our study reveals that the lift amplitude of the force fluctuations can be decreased by at least two-thirds through passive pitching. The efficacy of the unsteady load mitigation is only weakly dependent on the exact pitching axis location, and the optimal position is upstream and close to the axis of the foil. These results may inform the design of passive control systems of wind/tidal turbines and aerial/underwater vehicles and provide new insights into interpreting the control strategy of natural flyers such as insects and birds.

*Corresponding author
i.m.viola@ed.ac.uk (I.M. Viola)
www.voilab.eng.ed.ac.uk (I.M. Viola)
ORCID(s):

# 1. Introduction

Compared to large-scale aircraft, unmanned aerial vehicles (UAV) including small drones are highly susceptible to disturbances such as gusts and atmospheric turbulence (Findeis et al., 2019; Di Luca et al., 2020; Floreano and Wood, 2015), bringing challenges to their control, flight stability, efficiency and resilience (Watkins et al., 2006; Kambushev et al., 2019; Jones et al., 2022). With the increased use of drones in urban environments and in extreme weather conditions, improving the efficiency of control systems that mitigate unsteady load fluctuations is paramount.

While unsteady load mitigation is critical for UAVs, it is also important in a wide range of engineering applications such as dynamic positioning and navigation of autonomous underwater vehicles (Paull et al., 2014; Aguiar and Pascoal, 2007; Leonard and Bahr, 2016; Leonard and Graver, 2001; Anderson and Chhabra, 2002), vibration and noise of aeroplanes and helicopters (Seifert et al., 2004; Stanewsky, 2001; Iii et al., 2011; Kuder et al., 2013; Afonso et al., 2017; Li et al., 2018), and energy harvesting efficiency and survivability of wind and hydrokinetic turbines (Barlas and van Kuik, 2010b; Andersen, 2005; MacPhee and Beyene, 2015). Though active systems are broadly applied because they offer precise and adaptable control over dynamic conditions, such as active pitch systems for rotor blades (Sarkar et al. (2020); Barlas and van Kuik (2010a)), these are unable to suppress high-frequency unsteady loadings due to the delayed response. In fact, the faster the desired response, the higher the power requirement of the active system. Furthermore, most of active control systems add significant complexity to the overall system resulting in inherent lower reliability, which is undesirable as cost savings are substantially affected by system reliability (Njiri and Söffker (2016)). Lift control surfaces such as trailing edge flaps are typically actively controlled (Lackner and van Kuik, 2010) rather than set to respond to load fluctuations passively (Arredondo-galeana et al., 2021), but the latter approach is increasingly of interest because of the inherent higher reliability and zero energy consumption (Cognet et al., 2017; Arredondo-galeana et al., 2021). However, challenges also exist for exisitng passive systems. While simpler and potentially more robust, existing passive systems are usually designed to respond to a specific set of conditions, which can limit their effectiveness in highly variable environments (Greenblatt and Williams (2022)).

Birds' and insects' passive musculoskeletal response is often an inspiration for the design of the kinematic response of lifting surfaces (Hoerner et al., 2021; Stowers and Lentink, 2015; Chen et al., 2019), as their ability to fly stably in turbulence is superior to that of man-made vehicles. While gliding and hovering, they adopt a combination of active and passive systems to control their position and velocity, and to mitigate the effect of gusts and lulls (Nishikawa et al., 2007; Harvey et al., 2019, 2022; Bergou et al., 2010; Vance et al., 2013; Mistick et al., 2016). For example, the kestrel's (*Falco tinnunculus*) ability to keep its eye in a fixed position with respect to an earth-fixed frame while hovering has raised great curiosity from both researchers and bird watchers (Videler et al., 1983; Videler and Groenewold, 1991; Meyers, 1992; Fisher et al., 2015). Dissecting active neurological control and passive musculoskeletal responses is a major challenge, but recent discoveries suggest that passive musculoskeletal response is widely adopted for unsteady load mitigation (Cheney et al., 2020, 2014; Harvey and Inman, 2022; Ravi et al., 2016; Dakin et al., 2018). However, the extent to which load fluctuations can be mitigated through passive engineering systems or musculoskeletal responses is yet to be fully understood. This is partially due to the large number of parameters that govern the problem, including the wide range of wing shapes, gust profiles, and passive pitch kinematics.

Despite the large number of parameters, some general conclusions can be drawn by considering a simplified model that isolates the fundamental physics and neglects the details of the different applications. We do this by considering a two-dimensional (2D) rigid foil in an incompressible flow. The initial flow velocity is uniform and constant, and therefore exerts a constant torque on the foil, which is held in position by an externally-applied constant torque around the centre of rotation, hereafter the pitching axis. The velocity then varies over a transitory period to a different uniform velocity, and the foil passively moves to a new equilibrium position. In a biological system, the externally applied torque represents the musculoskeletal tension exerted on a joint. In an engineering system, the torque could be provided by a torsional spring.

This model was previously considered by Viola et al. (2021), who investigated analytically the effect of quasi-steady variations of both the freestream speed and direction. The authors showed that the chordnormal force component, and thus the lift for a small angle of attack, is kept constant. This occurs for low spring stiffness and high preload, such that the torque is constant and independent of the angular position of the plate. Their analysis suggests that passive pitch can be effectively used for unsteady load mitigation, but their quasi-steady analysis applies only to slow flow fluctuations. The present numerical investigation complements the theoretical analysis of Viola et al. (2021) by considering fast and large amplitude changes in the free stream speed and direction.

The engineering relevance of this conceptual passive pitching system is showed, for example, by its application to passive pitch systems for rotor blades. Pisetta et al. (2022), Dai et al. (2022), and Gambuzza et al. (2023b) tested analytically, numerically and experimentally, respectively, a turbine rotor with blades connected at the root through a bearing and a torsional spring. When the blade passes through a shear flow stream, it experiences a once-per-revolution variation in the onset flow speed and direction. The three papers report a substantial reduction in the thrust and torque fluctuations for the passively pitching blades compared to the fixed-pitch blades. Similarly, a passive pitch propeller blade was developed by Heinzen et al. (2015), who demonstrated a large expansion in the range of efficient operational conditions compared to a fixed pitch propeller.

It is noted that the combination of a rotation and a translation can be defined as a single rotation around a centre of rotation. Hence, while the present work may be considered to include both rotations and translations, we limit our study to fixed centres of rotation over the entire dynamic response. This means, for example, that the foil cannot first translate and then rotate. Furthermore, we note that any in-plane translation of a rigid body is equivalent to a pure rotation around a centre infinitely far from the body. For example, a horizontal translation is equivalent to a pure rotation around a centre at an infinite distance above or below the foil. Similarly, a vertical translation of the foil is equivalent to a pure rotation around a centre at an infinite distance left or right of the foil. However, a pure translation is not a physical solution because the foil must change angle of attack to reach a final equilibrium position.

The results of this work pave the way for interpreting biologically optimised solutions in natural flyers and for designing more efficient and reliable systems for unsteady load mitigation. We intentionally do not consider specific applications, as our findings are of a general nature and can be implemented to a wide range of possible design solutions. Overall, the present work identifies a broad range of pitching axis locations that can mitigate the lift amplitude of the force fluctuations by at least two-thirds through passive pitching. We also demonstrate that the efficacy of load mitigation is only weakly dependent on the exact pitching axis location, which lays the groundwork for practical engineering solutions. For example, we envisage that our work could inform the optimal design of passive pitch systems for wind and tidal turbines, such as the systems recently demonstrated by Pisetta et al. (2022) and Gambuzza et al. (2023b).

The rest of the paper is organised as follows. First, we describe the numerical model (§2.1) and we discuss its uncertainty (§2.2). Subsequently, we present the results of the numerical simulations (§3). We discuss separately the dynamic response to a fast change in the flow speed (§3.1) and direction (§3.2), then we study the optimal pitching position to minimise the load fluctuation (§3.3) and, finally, we extend the results to higher Reynolds numbers (§3.4). The paper closes with a summary of the main results (4).

## 2. Method
### 2.1. Numerical model

We model a 2D NACA0012 foil with a weakly coupled fluid-structure interaction (FSI) approach, based on computational fluid dynamics (CFD) implemented in the open-source toolbox `OpenFOAM` v2106. The initial angle of attack $\alpha_0 = 5°$ is chosen because it is an intermediate angle between the zero-lift condition for a symmetric foil and the stall angle, which is typically of the order of 10 degrees.

We have tested a range of the initial Reynolds numbers ($Re_0 \equiv u_0 c/\nu$, where $c$ is the chord length, $u_0$ is the initial inflow speed, and $\nu$ is the kinematic viscosity of the fluid) from $Re_0 = 10^3$ to $Re_0 = 10^6$, as this range covers various regimes from natural flyers (insects and small birds) and nano/micro air vehicles to large air vehicles and wind/tidal turbines. We primarily focus on $Re_0 = 10^3$, because it is sufficiently high to ensure the vorticity diffusion is negligible compared to advection, and sufficiently low to suppress lift fluctuations for a constant onset flow and angle of attack.

A rectangular computational domain (Fig. 1), is discretised with a structured mesh of hexahedral elements. The mesh is built with ICEM-CFD. Uniform Dirichlet conditions are enforced on the velocity and pressure on the two upstream and downstream boundaries of the domain, respectively. The velocity-pressure coupled equations are solved with the `pimpleFoam` solver using a Gauss linear spatial discretisation scheme and an Euler implicit time scheme with a Courant number lower than one.

For simulations where the initial Reynolds number is $10^3$, no turbulence model is applied. For $Re_0 = 5 \times 10^4$ and $Re_0 = 10^6$, the SST $k - \omega$ turbulence model is used with $y^+ < 1$. At every time step, the computed fluid torque on the foil is the input to the structural solver (`sixDoFRigidBodyMotion`), which solves Euler's second equation of motion with a second-order implicit Newmark scheme. The foil has a density of six times that of the fluid. The computed foil displacement is then used as an input to the morphing mesh solver (`dynamicMotionSolverFvMesh`), which morphs the mesh to fit the new position of the foil in the computational domain. The employed FSI approach is validated by modelling a benchmark case of an oscillating cylinder and comparing it to existing literature (see Appendix A).

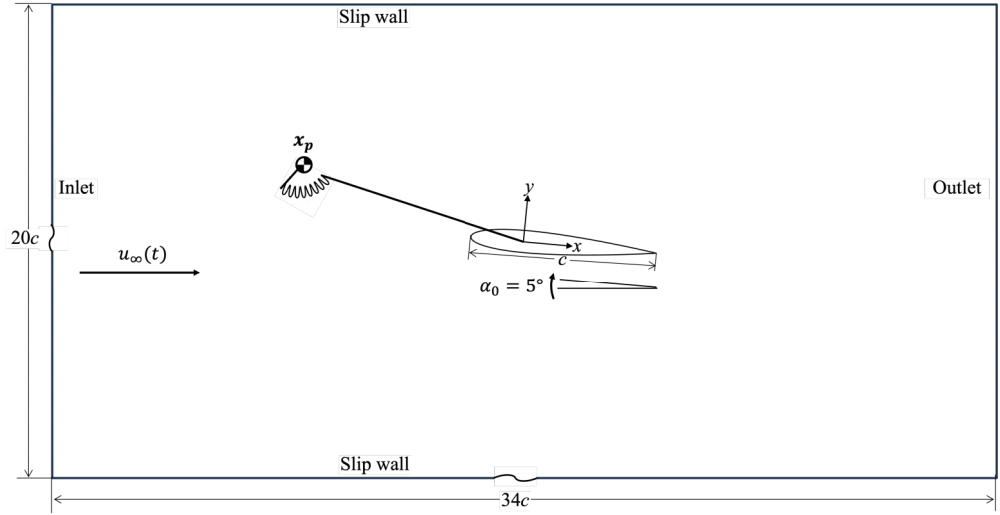

**Fig. 1:** Schematic of the computational domain and the passive pitch system. The *x* coordinate is along the foil chordwise direction, and the origin is placed at the aerodynamic center. *c* denotes the foil chord length.

## 2.2. Uncertainty quantification

We consider the uncertainty of the fluid forces computed with the CFD simulations following the least square approach proposed by Viola et al. (2013), which is summarised here. The verification process is set out to estimate the absolute uncertainty $E_\Phi$ at 95% confidence level in the estimate of the computed quantity $\Phi$ (i.e. the lift, the drag, etc.) through the estimate of the relative numerical uncertainty $U_\Phi$, where

$$E_\Phi = \Phi U_\Phi. \tag{1}$$

The relative numerical uncertainty is estimated independently for each source of error, including that due to the finite grid resolution $U_{\Phi_g}$, the finite time step ($U_{\Phi_t}$), the round-off ($U_{\Phi_r}$), and the finite number of iterations ($U_{\Phi_c}$). The uncertainties due to the finite spatial and time resolutions are estimated by undertaking simulations with different resolutions. Take $h$ as the relative step size of the resolution, i.e. the ratio between the linear grid size of the current and the reference grid or the ratio between the current and the reference time step. We define

$$\varphi(h) \equiv \frac{\Phi(h)}{\Phi_{\text{base}}}, \tag{2}$$

where $\Phi_{\text{base}}$ is the value of the computed solution with the base grid size and time step for which the uncertainty is computed. We then fit $\varphi(h)$ with

$$\varphi(h) \approx \zeta h^\xi + \varphi_0. \tag{3}$$

The coefficients $\zeta, \xi$ and $\varphi_0$ are exactly determined when $\varphi(h)$ is known for exactly three values of $h$; otherwise, they are computed with a least-squares method and $\sigma$ is the standard deviation of the fit.

The relative numerical uncertainty due to the spatial or time resolution ($U_{\Phi_h} = U_{\Phi_g}$ or $U_{\Phi_t}$) is

$$U_{\Phi_h} = \begin{cases} 1.25 \mid 1 - \varphi_0 \mid + \sigma & p \geq 0.95, \\ 1.5 \frac{\varphi_{\max} - \varphi_{\min}}{1 - \frac{h_{\min}}{h_{\max}}} + \sigma & p < 0.95. \end{cases} \tag{4}$$

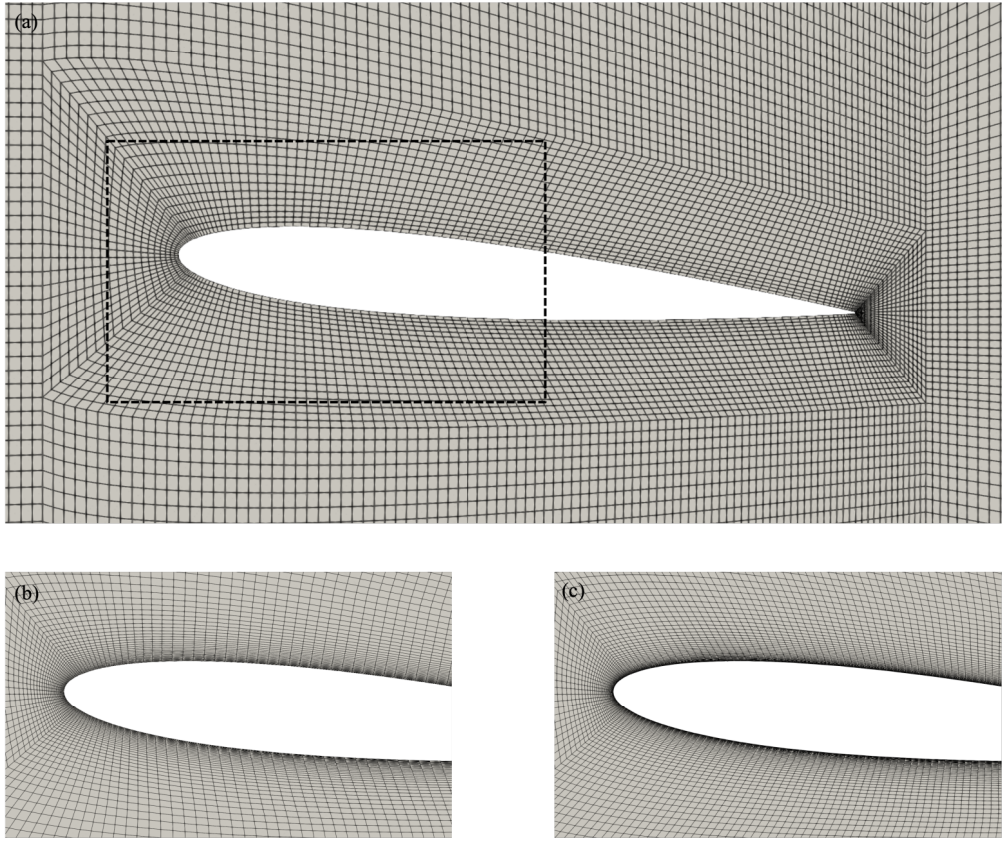

**Fig. 2:** Mesh around the foil for (a) $Re_0 = 10^3$ and contour (dashed box) of the zoomed-in view used for the higher $Re$ values, specifically: (b) $Re_0 = 5 \times 10^4$, and (c) $Re_0 = 10^6$.

where $\varphi_{max}$ and $\varphi_{min}$ are the maximum and the minimum of $\varphi(h)$, whilst $h_{max}$ and $h_{min}$ are the maximum and the minimum of $h$. The total relative numerical uncertainty is

$$U_\Phi = \sqrt{U_{\Phi_g}^2 + U_{\Phi_t}^2 + U_{\Phi_r}^2 + U_{\Phi_p}^2} + U_{\Phi_c}. \tag{5}$$

It is noted that $U_{\Phi_c}$ is not under the square root because it is not considered independent from the other sources of error. In the present work, we employ a double-precision solver and ensure that all residuals decrease by at least three orders of magnitude within each time step, resulting in negligible $U_{\Phi_r}$ and $U_{\Phi_c}$.

The uncertainty quantification for the simulations performed in the present work is undertaken for a foil that can passively pitch around P at $\boldsymbol{x}_P/c = (-1.5, 0.25)$, with an initial angle of attack of 5°, free stream velocity such that $Re_0 = 10^3$. The foil experiences a streamwise gust where the velocity doubles, $u_1 = 2u_0$, as shown in Fig. 3. The onset flow velocity is

$$u_\infty(t) = \begin{cases} u_0 & tu_0/c \leq 1, \\[2mm] \dfrac{u_0 + u_1}{2} - \dfrac{u_0 - u_1}{2} \tanh\left(\dfrac{t - t_0 - \frac{t_G}{2}}{\eta t_G}\right) & 1 < tu_0/c < 2, \\[2mm] u_1 & tu_0/c \geq 2, \end{cases} \tag{6}$$

where $u_0$ is the freestream velocity before the gust, $u_1$ that after the gust; $t_0$ is the instant when the gust starts, and $t_1$ the instant when the gust ends; $t_G = t_1 - t_0$ is the gust period; $\eta$ denotes the steepness and is set as 0.1 in the present work.

The choice of using a hyperbolic-tangent gust profile is based on the existing literature (Chowdhury and Ringuette (2021); Shao et al. (2024); Badrya et al. (2022)). The hyperbolic-tangent profile effectively smooths the sharp edges at the beginning and end of the gust, which is crucial for achieving stable and convergent simulations, especially when dealing with rapidly changing velocities. The condition with $t_G u_0/c = 1$ corresponds to that presented in Fig. 3, and a video of the time history of the flow field is available in Movie S1.

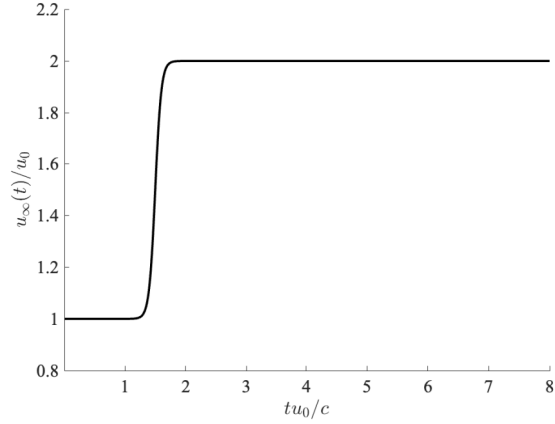

**Fig. 3:** Gust profiles for increasing free stream speed.

Simulations are performed at $Re_0 = 10^3$ on three geometrically similar grids: a coarse grid with 38,143 cells, a base grid with 72,368 cells, and a fine grid with 111,228 cells (Fig. 2). Fig. 4a and b show the fit of Eq. (3) where $\Phi$ is the dynamic peak of lift during the gust ($L_{max}$) and the final lift after the gust ($L_1$), respectively. The relative uncertainty due to the grid resolution $U_{\Phi_g}$ is found to be 0.34% and 1.38% for $L_{max}$ and $L_1$, respectively.

Simulations are performed with the base grid and four different time steps: $2 \times 10^{-4}$ $c/u_0$, $10^{-4}$ $c/u_0$, $5 \times 10^{-5}$ $c/u_0$ (base) and $10^{-5}$ $c/u_0$. Fig. 5a and b show the fit of Eq. (3) where $\Phi = L_{max}$ and $L_1$, respectively. The relative uncertainty due to the time resolution $U_{\Phi_t}$ is 0.027% and 0.61% for $L_{max}$ and $L_1$, respectively. The total relative numerical uncertainty $U_\Phi$ is computed with Eq. (5) as 0.34% and 1.51% for $L_{max}$ and $L_1$, respectively.

The same verification procedure is undertaken with finer grid and time resolutions at higher Reynolds numbers, resulting in a similar level of uncertainty for a grid size of 78,158 cells and a time step of $10^{-5} c/u_0$ at $Re_0 = 5 \times 10^4$, and 90,488 cells and a time step size of $10^{-5} c/u_0$ at $Re_0 = 10^6$.

As there are no experiments of the flow conditions here presented, the present FSI simulations have been validated by considering a cylinder free to oscillate both in the freestream and streamnormal direction. The validation is presented in the Appendix.

## 3. Results

The dynamic responses of the foil to changes in the free stream speed and direction are investigated separately in §3.1 and §3.2, respectively. In these two sections, the general gust response is discussed for an example pitching axis position. Additionally, the gust response for different positions of the pitching axis is presented in §3.3 and the results are extended to high Reynolds number flows in §3.4.

### 3.1. Dynamic response to fast-changing inflow speed

First, we consider fast changes in the flow speed at a constant angle of attack, i.e. streamwise gusts. Here we consider the extreme case of a sufficiently fast gust such that the gust response is independent of the duration of the transition from $u_0$ to $u_1$. Consider the same flow condition as that for which the uncertainty analysis was presented in §2.2, where the foil can passively pitch around P at $x_P/c = (-1.5, 0.25)$, with $Re_0 = 10^3$. The velocity doubles as described in Eq. (6). It must be noted that doubling the free stream velocity in less than a convective period is representative of an extreme gust.

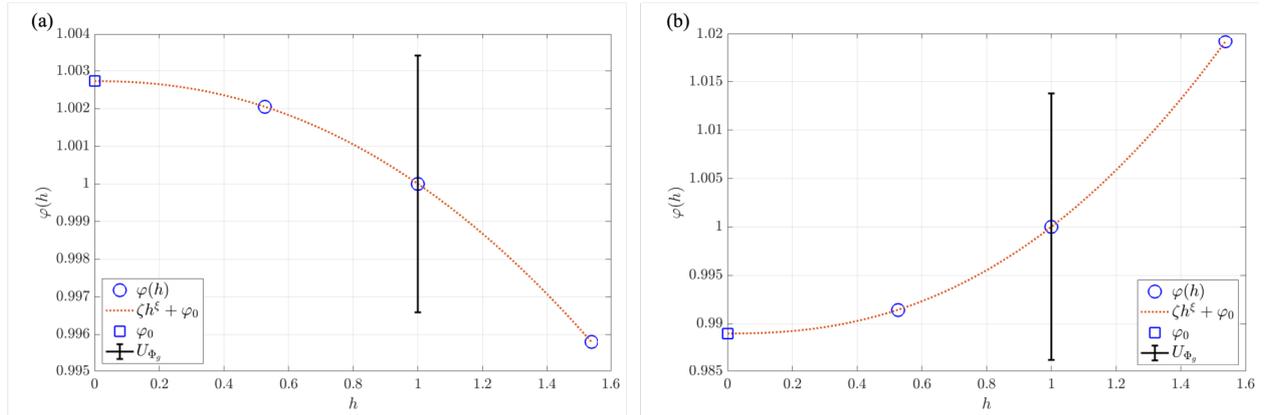

**Fig. 4:** Least square fit of Eq. (3) for the estimate of the relative uncertainty due to the grid resolution with (a) $\Phi = L_{\max}$, and (b) $\Phi = L_1$.

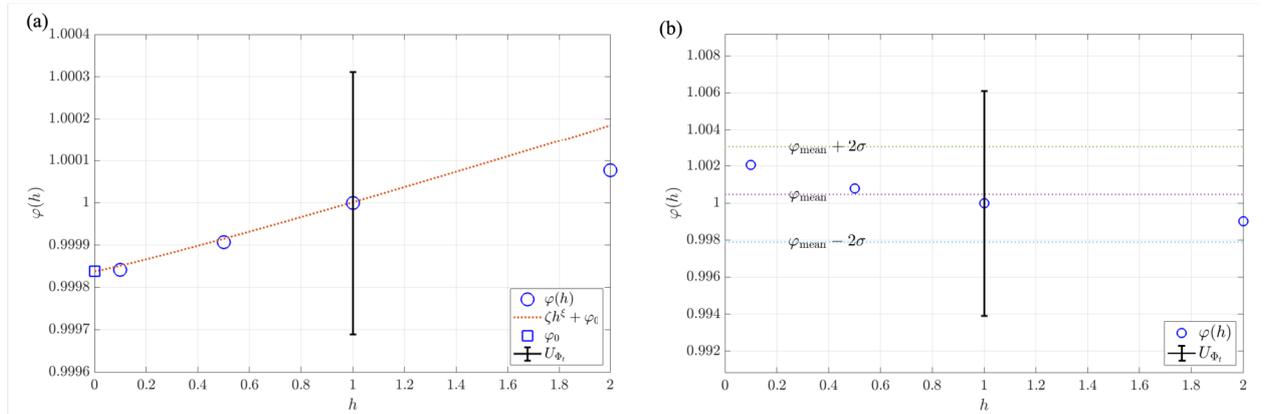

**Fig. 5:** Least square fit of Eq. (3) for the estimate of the relative uncertainty due to the time resolution with (a) $\Phi = L_{\max}$, and (b) $\Phi = L_1$.

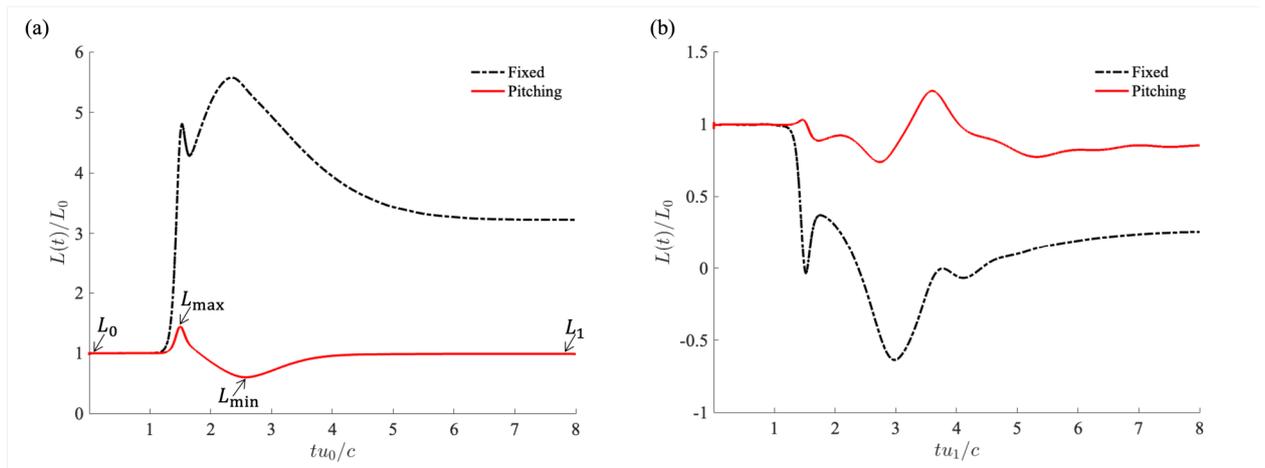

**Fig. 6:** Lift ratio versus nondimensional time for fixed and passively-pitching foils, with $Re_0 = 10^3$, $\mathbf{x}_P/c = (-1.5, 0.25)$ for (a) increasing inflow speed; (b) decreasing inflow speed.

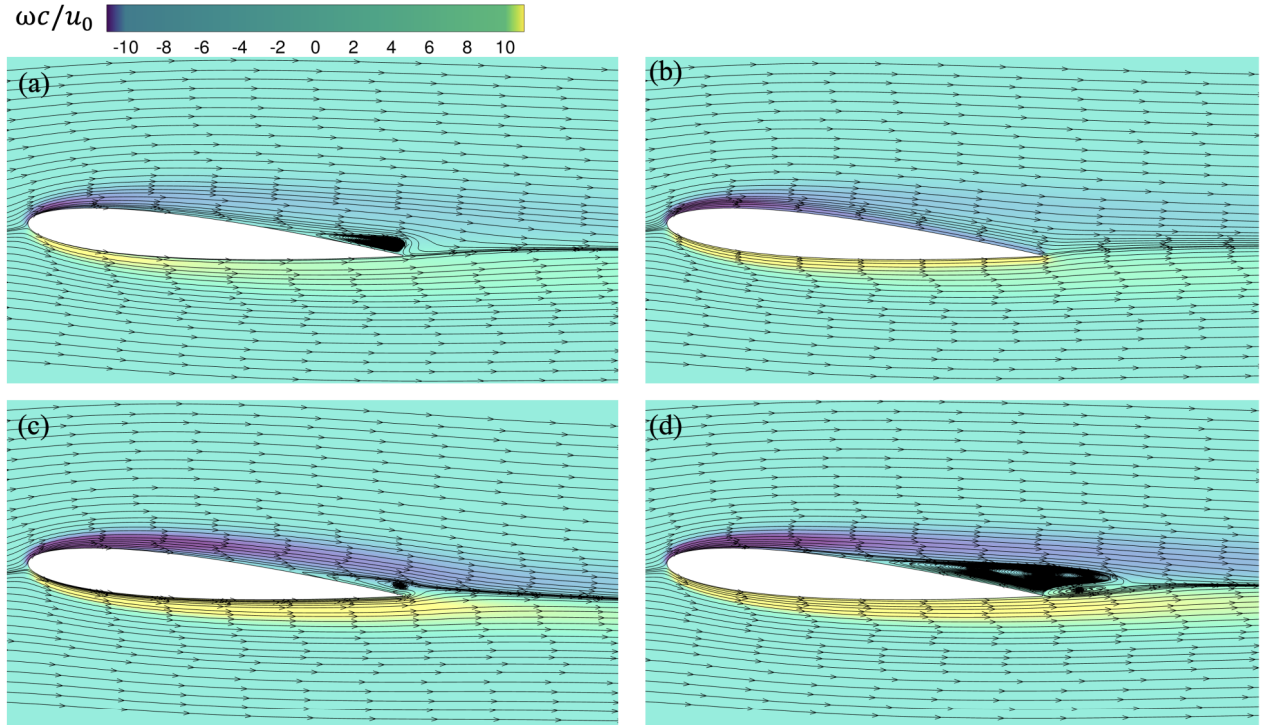

**Fig. 7:** Streamtraces and vorticity fields for the fixed foil experiencing increasing inflow speed, Eq. (6), from $Re_0 = 10^3$ with $x_P/c = (-1.5, 0.25)$: (a) $tu_0/c = 0.8$ (before the gust); (b) $tu_0/c = 1.5$ (the first peak, in middle of the gust); (c) $tu_0/c = 2.5$ (the second peak, after the gust); (d) $tu_0/c = 8.0$ (stable status, after the gust).

The time evolution of the ratio between the instantaneous lift $L(t)$ and the initial lift $L_0$ is shown in Fig. 6a. The initial condition shows a stable recirculation bubble near the trailing edge (Figure 7a). The fixed foil first experiences a sharp peak in lift due to the added mass, when the acceleration reaches the maximum at $tu_0/c = 1.5$. In the limit of an instantaneous change of the free stream velocity, a potential flow field is vectorially added to the existing flow field. The downstream stagnation point of this potential flow field is almost at the trailing edge, and thus a fast, strong gust, such as the one here considered, removes the recirculation region near the trailing edge (Figure 7b). After the added mass lift peak, the lift increases more progressively, following Wagner's function Wagner (1924).

At $tu_0/c \approx 2.5$, trailing edge separation occurs (Figure 7c), leading to an asymptotic drop in the lift (see Movie S1). The maximum lift is almost six times the initial lift. This is greater than predicted by Wagner's function (Wagner (1924)), because the initial lift coefficient (0.25) is much lower than $2\pi\alpha$ due to the presence of flow separation at the trailing edge. The final lift of the fixed foil is less than four times the initial lift because the trailing edge separation point moves upstream as the Reynolds number doubles (cf. Figure 7 and d).

In contrast, the passively pitching foil experiences a much lower lift fluctuation. First, the lift increases to a maximum value ($L_{max}$) due to the added mass, and this increase in lift causes the foil to pitch. Because of the inertia of the foil, the foil overshoots the equilibrium position, resulting in a minimum lift ($L_{min}$), which is lower than both the initial and final lift. For the passively pitching foil, the final lift is identical to the original lift.

Complementing Fig. 6a, where the free stream speed doubles, we investigate the dynamic lift response when the free stream speed halves. This is given by the Eq. (6), where $u_0 = 2u_1$ and the gust period is between $tu_1/c = 1$ and $tu_1/c = 2$.

As also shown by the comparison between Fig. 6a and Fig. 6b, halving the free stream speed results in larger amplitude load fluctuations than when the free stream speed doubles. The time history of the flow field for this condition is available in Movie S2.

The kinematics of the pitching foil is independent of $t_G$ for sufficiently low $t_G$. In fact, as $t_G \to 0$, the added mass force tends to infinity, and its duration tends to zero, whilst its impulse is constant. The effect of the added mass is to generate vorticity (with net zero circulation) on the surface of the foil with strength proportional to the added mass impulse. Hence, as long as the added mass vorticity is generated at the foil surface within a time scale $t_G$ that is much smaller than the advective time scale $c/u_0$, the vorticity field at the end of the gust is independent of $t_G$, and so is the kinematics of the pitching foil.

For a fixed foil, while the first added-mass lift peak observed in Fig. 6a tends to infinity as $t_G \to 0$, the second lift peak tends to a constant value. This is demonstrated by Fig. 8, which shows the amplitude of the lift fluctuation, excluding the first added-mass lift peak, versus the gust period $t_G$. The lift amplitude clearly shows an asymptotic trend both for $t_G \to 0$, and for $t_G \to \infty$, corresponding to the quasi-steady condition. In the following we consider $t_G u_0/c = 1$, for which the lift fluctuation is more than 98% of that extrapolated for $t_G \to 0$, i.e., the behaviour approaches that expected with an instantaneous velocity change.

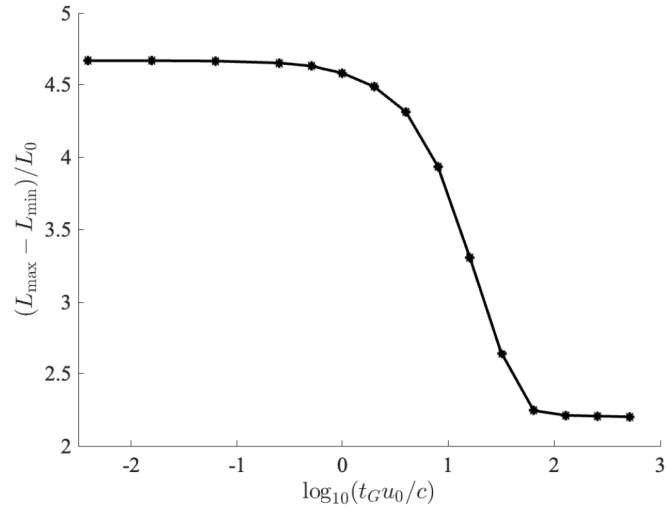

**Fig. 8:** Dynamic lift fluctuation versus the gust period $t_G$. Here the added mass peak is not considered, as it tends to infinity as the gust period tends to zero.

### 3.2. Dynamic response to fast-changing inflow direction

Gusts can involve both changing inflow speed and changing inflow direction. We now consider the response of a passively pitching foil to changes in angle of attack. Fig. 10a shows $L(t)/L_0$ when the angle of attack doubles from 5° to 10° between $tu_0/c = 1$ and $tu_0/c = 2$ (Fig. 9a). For the fixed foil, a first sharp peak due to the added mass lift is observed. Then, the lift gradually increases until $tu_0/c \approx 3$. Subsequently, the lift reduces due to the trailing edge separation (see Movie S3). Compared with the fixed foil, there is a significant decrease in the lift fluctuation experienced by the passively pitching foil.

Fig. 10b shows $L(t)/L_0$ when the angle of attack halves from 10° to 5° between the same period, following Eq. (7) with $\alpha_0 = 2\alpha_1$, as showed in Fig. 9b. An approximately reversed lift response, compared to the situation where the angle of attack doubles, is observed for both the fixed foil and the pitching foil. Again, passive pitching mitigates much of the lift fluctuation experienced by the fixed foil. The time history of the flow field for these two conditions is available in Movie S3 and Movie S4.

$$\alpha(t) = \begin{cases} \alpha_0 & tu_0/c \le 1, \\ \dfrac{\alpha_0 + \alpha_1}{2} - \dfrac{\alpha_0 - \alpha_1}{2} \tanh\left(\dfrac{t - t_0 - \frac{t_G}{2}}{\eta t_G}\right) & 1 < tu_0/c < 2, \\ \alpha_1 & tu_0/c \ge 2. \end{cases} \quad (7)$$

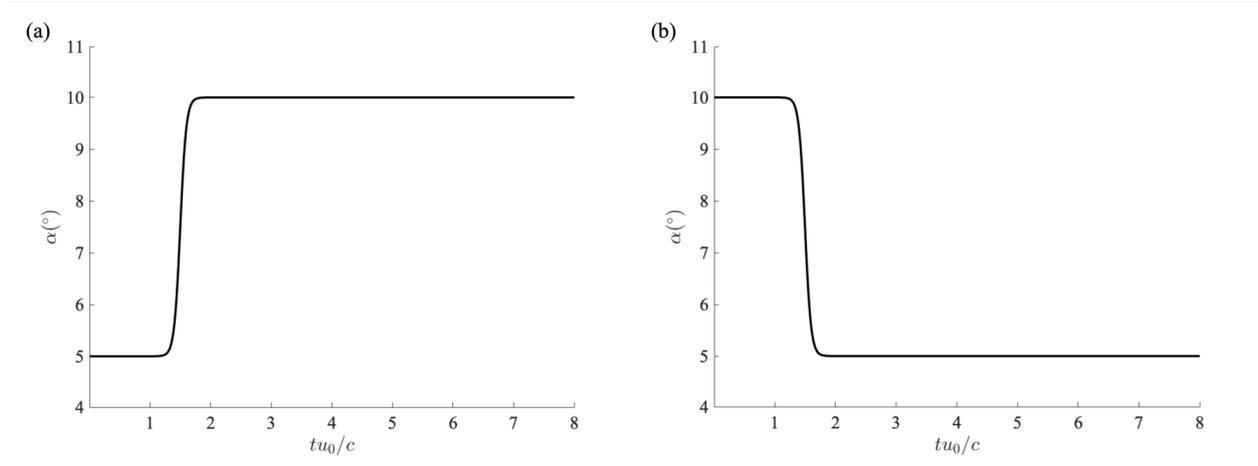

**Fig. 9:** Gust profiles for: (a) increasing free stream angle of attack; and (b) decreasing free stream angle of attack.

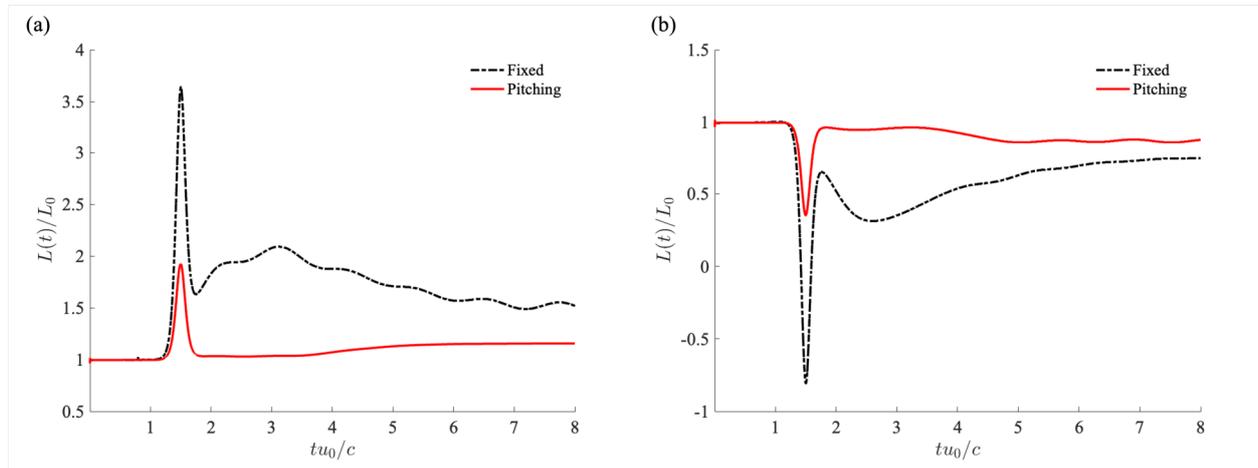

**Fig. 10:** Lift ratio versus nondimensional time for fixed and passively-pitching foils, with $Re_0 = 10^3$, $x_P/c = (-1.5, 0.25)$ and (a) increasing angle of attack; (b) decreasing angle of attack.

### 3.3. Optimal pitching axis position

So far, all the results presented have been for a pitching axis at $x_P/c = (-1.5, 0.25)$. In this section, we will consider the effect of moving the pitching axis relative to the foil. We quantify the dynamic unsteady load mitigation as the ratio between the amplitude of the lift variations for the passively pitching foil and the fixed foil:

$$\epsilon_{\text{DY}} = \frac{(L_{\max} - L_{\min})_{\text{pitching}}}{(L_{\max} - L_{\min})_{\text{fixed}}} \tag{8}$$

Fig. 11a shows contours of $\epsilon_{\text{DY}}$ for a range of pitching axis locations in front of and above the foil (i.e. $x_P < 0$ and $y_P > 0$), where the maximum unsteady load mitigation is found. The gust considered is the rapid speed increase (as in Fig. 3). In this example, the pitching position can be selected such that the pitching foil experiences less than 18% of the lift fluctuations experienced by the fixed foil.

Notably, a sudden drop in the freestream velocity is more difficult to mitigate than a velocity increase. This is shown in Fig. 11b, where the velocity halves between $tu_0/c = 1$ and 2. This result is not surprising: an instantaneous flow acceleration promotes flow reattachment on a foil with trailing edge separation, while a flow deceleration tends to shift the separation point upstream (Movie S2). Nevertheless, the simulations show that the passively pitching foil can be optimised to experience less than one-third of the lift fluctuation amplitude of a fixed foil (Fig. 11b). The optimal pitching axis is also located in front and above the foil, but the optimal chordnormal position $y_P$ is somewhat lower than that to mitigate a speed increase.

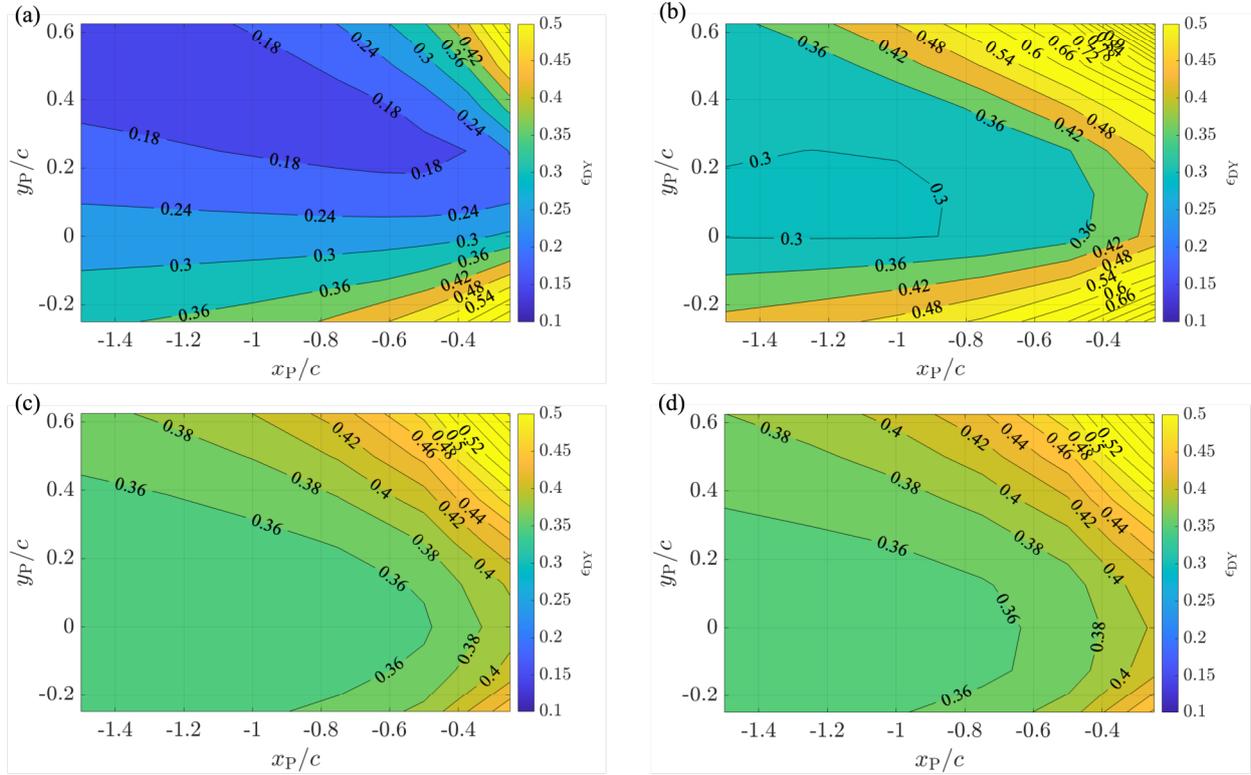

**Fig. 11:** Ratio of dynamic lift fluctuation $\epsilon_{DY}$ versus $x_P$ for the pitching foil at $Re = 10^3$ experiencing:(a) increasing inflow speed. (b) decreasing inflow speed. (c) inflow with increasing angle of attack. (d) inflow with decreasing angle of attack.

A similar level of unsteady load mitigation is achieved for a change in the freestream direction while the flow speed remains constant. In Fig. 11c, the initial angle of attack is $\alpha_0 = 5°$, and the freestream direction varies such as to increase the angle of attack of the fixed foil up to $10°$ within less than one convective period (Fig. 9a and Movie S3). Conversely, in Fig. 11d, the angle of attack of the fixed foil varies from $10°$ to $5°$ (Fig. 9a and Movie S4). The results reveal that the passively-pitching foil can also experience about 35% of the lift fluctuations of the fixed foil for changing inflow direction (Fig. 11c and Fig. 11d). Akin to the previous results, the loci of the optimal pitching axis is upstream of the foil, but rather than above the foil, it is along the chord line ($y_P = 0$).

Overall these results reveal that passive pitch is effective in mitigating the force fluctuations due to fast, large-amplitude changes in the free stream speed and direction. The load fluctuations due to a rapid change of $5°$ in the angle of attack are more difficult to mitigate than the sudden doubling or halving of the flow speed. Even in these extreme cases, however, about two-thirds of the fluctuations can be suppressed. As was seen for quasi-steady variations in free stream velocity, for mitigating fast gusts, the optimal locus of the pitching axis is along a line upstream of the foil.

### 3.4. High Reynolds number conditions

The simulations so far have been carried out under laminar flow conditions at $Re_0 = 10^3$. In this section, we consider a streamwise gust at $Re_0 = 5 \times 10^4$ and $Re_0 = 10^6$, with the turbulence models described in Section 2.1. Fig. 12 shows the time history of the lift ratio for a step increase in the flow speed following Eq. (6). The comparison between the fixed and the pitching foil allows us to draw similar conclusions to those drawn at $Re_0 = 10^3$ (Fig. 6a). The lift experienced by a pitching foil recovers to that before the gust, and the dynamic load fluctuations are significantly suppressed. At $Re_0 = 10^6$, the lift coefficient is not constant after the gust as the gust triggers an unsteady load oscillation. This is because boundary layer separation occurs, and there is a vortex shedding formed on the foil suction side (shown in Movie S6). But this is also effectively mitigated by the passive pitch mechanism, owing to the suppressed boundary layer separation at a reduced angle of attack, as shown in Movie S5 ($Re_0 = 5 \times 10^4$) and Movie S6 ($Re_0 = 10^6$).

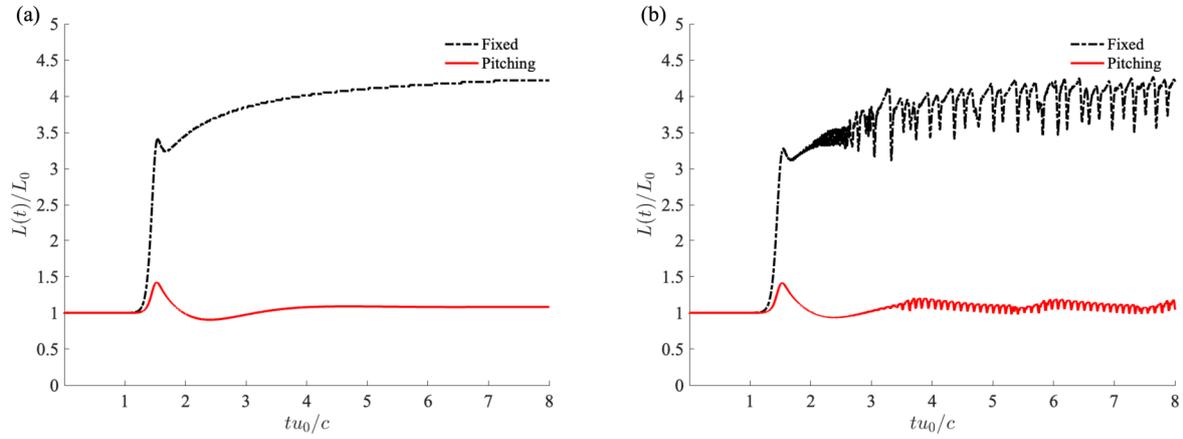

**Fig. 12:** Lift ratio versus nondimensional time for fixed and passively-pitching foils, with $x_P/c = (-1.5, 0.25)$, increasing inflow speed, (a) $Re_0 = 5 \times 10^4$, and (b) $Re_0 = 10^6$.

## 4. Conclusions

Load alleviation is of critical importance in a range of applications, from wind turbines to air taxis. So far, designers have favoured active load mitigation devices despite their inherent high cost and complexity, while passive systems have received comparatively little attention despite their common occurance in nature. This work considers the general case of a foil that is free to pitch about an axis and therefore elucidates general physical principles that are common to any application.

The present investigation uses fluid-structure interaction simulations to reveal that load fluctuations can be mitigated by over 80% by simple passive pitching of the foil. Therefore, the first result of this work is that passive mitigation of fluid load fluctuations is highly effective for both slow and fast changes in either the flow speed or direction. Specifically, even in the extreme gust conditions considered here, two-thirds of the load fluctuations are cancelled. This is possible by placing the pitching axis upstream of the foil, at $y_P/x_P$ between 0 and 0.3. The exact position depends on the foil geometry and inertia, as well as on the initial and final flow velocities.

The optimal location of the pitching axis depends slightly on the gust type. However, the sensitivity of the unsteady load mitigation to the exact position of the pitching axis is low in the region of the optimal position. Therefore, the precise position of the pitch axis is not important for unsteady load alleviation, and it is likely to be dictated by other biological or design constraints as long as it is located near the chord line upstream of the wing.

Although the resultant fluid force varies linearly with the angle of attack and quadratically with the flow speed, we found the mitigation of $\pm 5°$ angle of attack variations more difficult than mitigation of a twofold change in the flow speed. The sensitivity of the system efficacy to the amplitude of the angle of attack variations is the main limitation of the passive system.

Our findings provide significant insights into the potential of passive pitching to mitigate lift fluctuations across a wide range of flow conditions. By demonstrating that the efficacy of load mitigation is only weakly dependent on the exact pitching axis location in a broad range, our study lays the groundwork for practical engineering solutions. The present work will inform the design of passive engineering systems, which may be inherently more reliable than complex active systems. For instance, we expect these findings to be translated into design guidelines for various applications, such as wind/tidal turbines, aerial/underwater vehicles, and bio-inspired flying robots. For example, we have designed and tested a passive pitching system using a torsional spring and a mounting arm (Gambuzza et al. (2023a)), which has demonstrated a prominent effect on mitigating unsteady load fluctuations for tidal turbines. They may also contribute to gaining new insights into the passive musculoskeletal systems of biological flyers.

## 5. Acknowledgements

This work was supported by the UK Engineering and Physical Sciences Research Council through the grant 'Morphing-Blades: New-Concept Turbine Blades for Unsteady Load Mitigation' [EP/V009443/1]

## A. Appendix

A 2D oscillating cylinder, for which data is available in the literature Blackburn and Karniadakis (1993); Menon and Mittal (2019), is modelled to validate the FSI approach. The cylinder is immersed in a uniform and constant stream at the Reynolds number of 200 and is connected to a spring and a damper with one degree of freedom along the crossflow direction $y$. Note that the damping is introduced in these simulations of ours to match the validation case, while the damping is not considered in our study.

Fig. 13a shows the nondimensional cylinder oscillation amplitude versus the mass-damping parameter $\gamma \equiv 8\pi^2 \hat{\zeta} St^2 m/\rho D^2$ with $St = 0.1959$ and $m/\rho D^2 = 10$, where $\hat{\zeta}$ is the damping ratio, $\rho$ is the fluid density, $m$ is the cylinder mass, and $D$ is the cylinder diameter. The Strouhal number $St$ is based on the vortex shedding frequency around the fixed cylinder. Fig. 13b shows the flow-induced oscillation displacement phase plot for $\hat{\zeta} = 0.01$, $m/\rho D^2 = 1$. The calculated results agree well with the existing literature Menon and Mittal (2019); Blackburn and Karniadakis (1993).

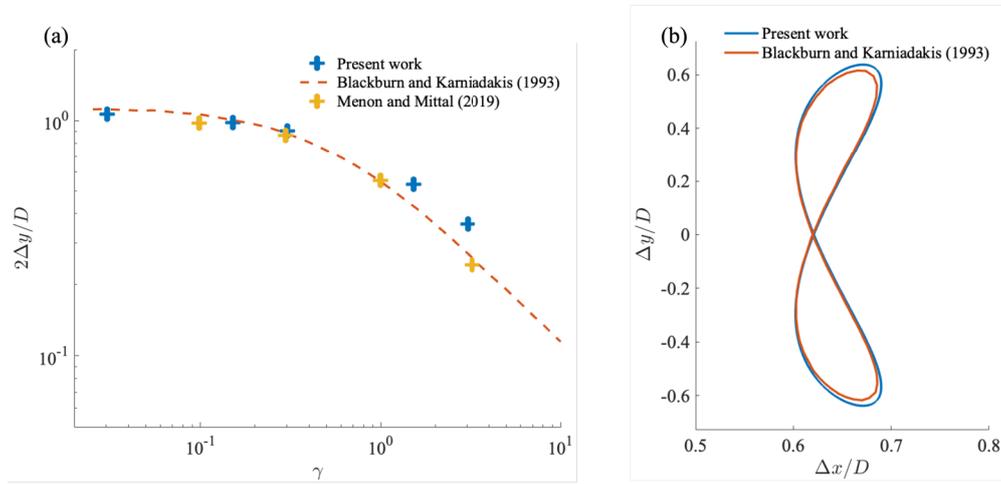

**Fig. 13:** (a) Flow-induced streamnormal oscillation amplitude of a cylinder versus the mass-damping parameter at $Re = 200$ and $\hat{\zeta} = 0.01$, and (b) streamnormal versus streamwise displacement for $\hat{\zeta} = 0.01$, $m/\rho D^2 = 1$. Data include present work and other authors Menon and Mittal (2019); Blackburn and Karniadakis (1993).

Fig. 14 shows the trend of the lift peak value versus the spring stiffness when the pitching axis is located at the leading edge, and an asymptotic trend is observed when the spring is decreasing. This demonstrates that when the stiffness is reduced to a certain threshold, the lift fluctuation amplitude is independent of the further decrease of the spring stiffness. Therefore, we choose the nondimensional stiffness of $\kappa = 0.1018$ to exclude the influence of stiffness.

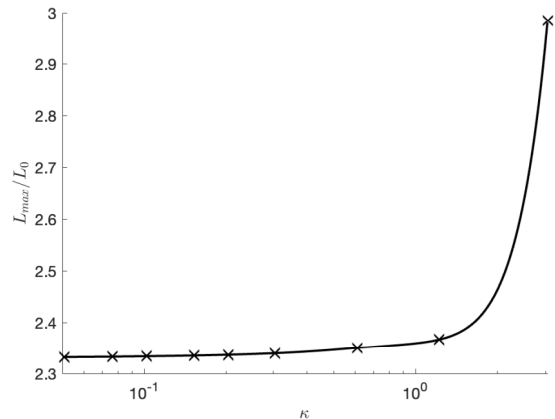

**Fig. 14**: Independence validation of stiffness. The spring stiffness $\kappa$ is nondimensionalised by $\frac{1}{2}\rho c/U_0^2$.